# Research on Education Big Data for Student's Academic Performance Analysis based on Machine Learning

Chun wang

LaFetra College of Education, University of La Verne, CA, USA, chun.wang@laverne.edu

Jiexiao Chen

Steinhardt School of Culture, Education, and Human Development, New York University, New York, USAic11181@nyu.edu

Ziyang Xie

Departmen of Art and Design, Hunan University of Humanities, Science and Technology, Hunan, ChinaXieziyang197843@gmail.com

Jianke Zou*

HSBC Business School, Department of Management, Peking University, Peking, China, Zoujianke@pku.org.cn

The application of the Internet in the field of education is becoming more and more popular, and a large amount of educational data is generated in the process. How to effectively use these data has always been a key issue in the field of educational data mining. In this work, a machine learning model based on Long Short-Term Memory Network (LSTM) was used to conduct an in-depth analysis of educational big data to evaluate student performance. The LSTM model efficiently processes time series data, allowing us to capture time-dependent and long-term trends in students' learning activities. This approach is particularly useful for analyzing student progress, engagement, and other behavioral patterns to support personalized education. In an experimental analysis, we verified the effectiveness of the deep learning method in predicting student performance by comparing the performance of different models. Strict cross-validation techniques are used to ensure the accuracy and generalization of experimental results.

## 1 INTRODUCTION

According to the definition of big data, big data refers to a data set that is too large to be analyzed and processed using conventional software. Educational big data refers to the data generated by students' daily learning and life activities from the school's academic affairs system, financial system, library system and other sources. The research on education big data mainly includes employment analysis, student specialty analysis, student development trajectory, online behavior analysis, and student portrait [1]. By analyzing this data, it is possible to make predictions about student behavior and potential risks, laying the foundation for providing personalized educational services.

Many higher education students face serious academic challenges, including weak self-control and a lack of motivation for active learning, which often require external supervision and intervention. In addition, some students are at risk of dropping out and having difficulty obtaining a degree due to the large number of failed subjects, and some students are under scrutiny for violations such as late arrivals, early departures, and absenteeism. Internet addiction, in particular, makes it difficult for some students to meet academic requirements [2]. These problems not only lead to increased psychological stress for some students, but also lead to self-harm behavior. In order to reduce these phenomena, schools and relevant authorities need to take measures to strengthen the supervision of students' learning, intervention in inappropriate behavior, and counseling for students with psychological problems.

In the face of these challenges, human resources often appear to be insufficient. Therefore, it is crucial to use educational data mining to analyze and predict students' academic performance. With the increase of campus service management platforms, the accumulation of student data has shown massive growth, including consumption data, living habits and learning conditions. By making efficient use of this data, as well as applying machine learning and deep learning techniques, it is possible to better analyze and evaluate students' academic lives, as well as predict academic performance, so as to effectively intervene early on students who are underperforming [3]. This not only improves students' academic performance, but also helps them have a more meaningful school life.

Students are at the center of educational and teaching activities, and they are also the focus of school management and educators. Using big data in education, it is possible to analyze students' behavior patterns and lifestyle habits, and then explore how these behaviors affect academic performance. Studies have shown that students' daily behavior patterns are positively correlated with academic performance. In addition, good online habits generally reflect greater self-control, which is also positively correlated with academic performance. At the same time, by analyzing students' behavior on social media, the researchers found that social media activity is closely related to students' mental health [4]. Effective use of big data in education can not only identify students who may be at risk of academic performance in a timely manner, but also take targeted measures to improve students' academic performance and mental health.

Through education data mining, valuable information can be extracted from a large amount of data on educational activities, which supports education administrations and educators in policy and decision-making. The core of educational data mining is to extract useful higher-order features from the vast and complex raw data, and use the actual data records of students for analysis [5]. Information on students' courses and families, collected through questionnaires and other means, is also included in the analysis to predict students' future performance and potential, which is particularly important for higher education institutions in the student selection process.

This method not only helps to understand a student's past performance, but also predicts their future entrance test scores and aptitude to select the most suitable candidates [6].

Evaluating students' academic performance is a critical part of the educational process. By using existing student information to predict their future performance, it can effectively reduce the failure rate, improve the quality of teaching, and optimize students' academic performance. With the widespread promotion of higher education and the deepening of educational research, the impact of the teaching environment on students' academic performance has become increasingly obvious. At present, research on academic performance mainly focuses on using existing data to predict students' grades in each semester and whether they will fail courses. This approach not only helps educators better understand students' learning needs, but also provides data to support the development of more effective instructional strategies.

## 2  RELATED WORK

Recently, researchers Aya et al. [7] conducted a study on the prediction of students' successful completion of courses such as "programming" and "data structures", which are easy for students to fail. They used a publicly available student achievement dataset at a four-year university to compare the performance of deep neural networks (DNNs), random forests, support vector machines (SVMs), and K-nearest neighbors (KNN) models. The results show that the deep neural network has the best prediction effect, with an accuracy rate of 89%. This further demonstrates the effectiveness of deep neural networks in predicting student academic performance. In addition, Hyoungjoon et al. [8] focused on students' location information, extracting information from students' appointment records and card swiping records, classifying locations into four categories and sorting out students' activity tracks. The dynamic time curvature (DTW) method was used to analyze the similarity of students' trajectories in one semester, and the bottom-up hierarchical clustering method was used to divide the trajectories into three categories, and the average GPA of each class was calculated. Their seven sets of trials showed a clear correlation between students' trajectories and their average GPAs.

Predicting students' grades at the beginning of the semester has always been an important topic in education big data research. Through education data mining, analyzing the correlation between students' data can effectively assist the education department in daily management and target support for students who may be at risk, which is extremely valuable for students, teachers and educational institutions. Sadiq Hussain et al. [9] applied deep learning regression models and linear regression models (RM) to educational datasets and solved the overfitting problem by adjusting parameters, and their studies showed that deep neural network models were significantly better than linear regression models in mining and predicting students' academic performance. On the other hand, Shahab Boum et al. [10] used the hidden Markov model to propose a standard and intuitive classification method to describe the level of students' academic performance, and constructed a compact learning performance trajectory to further explore the association between learning performance trajectory and academic success.

In recent years, the online teaching model has developed rapidly, and a large number of students' online learning information has been recorded in detail. Javier Bravo-Agapito et al. [11] constructed a series of models to predict early student achievement using Moodle interaction data and performance characteristics from 802 undergraduate students at CO University. These models included exploratory factor analysis, multiple linear regression, cluster analysis, and correlation analysis, and the four factors that were found to have the greatest impact on student achievement were: the number of visits to forums and glossaries, questionnaires, assignments, and student age. However, the excessive use of smartphones often has a negative impact on student achievement. Researchers Eve Sarah Troll [12] explored the relationship between smartphone use and student achievement, including duration of use, beneficial usage habits, and smartphone procrastination. Studies have shown that students with good smartphone habits tend to have greater self-control, which helps them achieve better results academically.

## 3 METHODOLOGIES

### 3.1 Notions

Following Table 1 summarizes commonly used notations.

Table 1: Notions

| Symbols | Definition |
| --- | --- |
| Title_document | main title of article |
| Subtitle | subtitle of article |
| Authors | author name |
| Affiliation | author affiliation information |
| ComputerCode | Display Computer codes |
| Short Title | Short title of article |

### 3.2 Long-short Term Memory Network

In order to further integrate education big data and LSTM models, we need to extend the basic architecture of LSTM to more effectively handle and analyze complex patterns and dependencies in education data. The purpose of the forgetting gate is to decide what information to discard from the cell state. It is calculated by the following Equation 1. Student learning activity data, login frequency, and assignment submission time can be input into the LSTM as a time series, and the model can learn the time dependence of student behavior to predict students' future performance or risk.

$$f_t = \sigma(W_f \cdot [h_{t-1}, x_t] + b_f) \#(1)$$

Where $\sigma(\cdot)$ is a sigmoid activation function that controls the circulation of information. $W_f$ is the weight of the forgetting gate, $b_f$ is the offset term, and $h_{t-1}$ is the hidden state of the previous time step. $x_t$ is the behavioral characteristic of the student at the current time step.

Non-time series features in education data, such as student demographics, can be fused into the model through extended input gates, increasing the prediction accuracy of the model. In order to integrate non-time series features in education data, these features can be incorporated by modifying the input gates. z is a non-time series feature, then the update formula for the input gate can be rewritten as Equation 2 and 3.

$$i_t = \sigma(W_i \cdot [h_{t-1}, x_t, z_t] + b_i) \#(2)$$
$$\tilde{c}_t = tanh(W_c \cdot [h_{t-1}, x_t, z_t] + b_c) \#(3)$$

By allowing the LSTM to take into account both time series and static features when updating the cell state, the accuracy of the prediction is improved.

In multi-task learning scenarios, LSTMs can be trained to predict multiple outputs simultaneously, such as student achievement, engagement, and dropout risk. For multi-task learning, the output layer can be designed as multiple independent units, each corresponding to a specific prediction task, which can be calculated as following Equation 4.

$$y_{c,GPA} = \sigma(W_{Grade} \cdot h_t + b_{Grade})$$
$$y_{c,Engagement} = \sigma(W_{Engagement} \cdot h_t + b_{Engagement}) \#(4)$$
$$y_{c,Risk} = \sigma(W_{Risk} \cdot h_t + b_{Risk})$$

### 3.3 Dynamical Adjustment

Dynamic adaptation is an important feature of the LSTM model, especially in response to the changing environment in the education system, such as the start of a new semester, the update of course content, or the change of teaching strategies. By dynamically adjusting the model parameters, the model can be more flexible to adapt to these changes, thus maintaining the accuracy and validity of its predictions. Dynamic adjustment is mainly achieved by modifying the learning rate of the model α or directly adjusting the model parameter W. This adjustment is based on the following update rules, which is shown in Equation 5.

$$W_n = W_o - \alpha \cdot \nabla_W L \#(5)$$

Where $W_n$ is the updated parameter and $W_o$ is the parameter before the update. Parameter $\alpha$ is the learning rate, which controls the step size at which the parameter is updated. Symbol $L$ is the loss function that evaluates the error between the model's prediction and the actual result. Calculation results of $\nabla_W L$ is the gradient of the loss function with respect to the parameter $W$, indicating the direction in which the loss function increases the fastest in the parameter space.

In education big data, dynamic adjustment can help the model adapt to changes in students' behavior patterns, such as different course loads in the new semester and changes in students' learning habits. For example, if a semester's course relies more on online sources than before, the model can adjust the parameters to give more weight to these new data patterns. Dynamic tuning not only improves the flexibility and adaptability of the model, but also helps maintain its performance over long runs, especially when dealing with non-static data sources. This is critical for education policymakers, who need to respond based on the most up-to-date data.

By adjusting the learning rate, we can control the speed at which the parameters are updated during model training, which affects the stability and convergence speed of training. Learning rate decay is a common strategy to gradually reduce the learning rate during the training process to help the model converge quickly in the early stage of training, and reduce the update step size when it is close to the optimal solution, so as to avoid excessive swing or even deviation from the optimal solution. Among them, exponential decay is a widely used attenuation method. The mathematical expression for exponential decay is as Equation 6.

$$\alpha = \alpha_o e^{-kt} \#(6)$$

Where $\alpha_o$ is the initial learning rate, which is usually set before the start of model training, and an appropriate value is selected based on the specific problem and network structure. $k$ is the decay constant, which determines the rate at which the learning rate decreases. The higher the $k$-value, the faster the learning rate decays. $t$ is the number of iterations, which is the number of steps or rounds in the training process of the model.

This approach can help the model jump beyond the local minimum and explore a broader parameter space, so that it is possible to find a better global minimum. Periodicity tuning usually relies on heuristic rules to set the upper and lower bounds of the cycle length and learning rate. Through a reasonable learning rate adjustment strategy, we can effectively control the learning dynamics in the process of model training, improve the stability and convergence of the model, and improve the accuracy of prediction and the versatility of the model when processing educational big data.

## 4 EXPERIMENTS

### 4.1 Experimental Setups

EdNet-KT1 is a large-scale education dataset designed for knowledge tracking, which integrates data on the interaction of millions of students on online learning platforms. This dataset includes detailed records of student activity, such as clickstream, time of answers, correctness of answers, and more, and is designed to support researchers and developers in using machine learning models to analyze and predict student learning behavior and performance. EdNet-KT1 is data-rich and multi-dimensional, making it suitable for in-depth research on personalized learning, learning path optimization, and other areas related to educational technology. In the experimental setup using the EdNet-KT1 dataset, the goal was to use the LSTM model to track and predict student learning performance.

The experiment first involves data preprocessing, including data cleaning and feature selection, and the selection of key features such as correctness of answers and timestamps. Specifically, the model parameters are set to 128 LSTM cells, and a dropout rate of 0.5 to reduce overfitting. The training uses the Adam optimizer, the initial learning rate is set to 0.001, the batch size is 32 or 64, and the total training period is 50. Performance evaluation is performed through metrics such as accuracy, precision, recall, and F1 score, while K-fold cross-validation is used to ensure the reliability of the results. This comprehensive experimental design aims to provide an in-depth analysis of student behavior patterns in online learning environments and to evaluate the effectiveness of instructional strategies.

### 4.2 Experimental Setups

Accuracy is the most intuitive performance metric, and it represents the proportion of the total number of samples that the model predicts correctly. Following Figure 1 shows the students' performance prediction results.

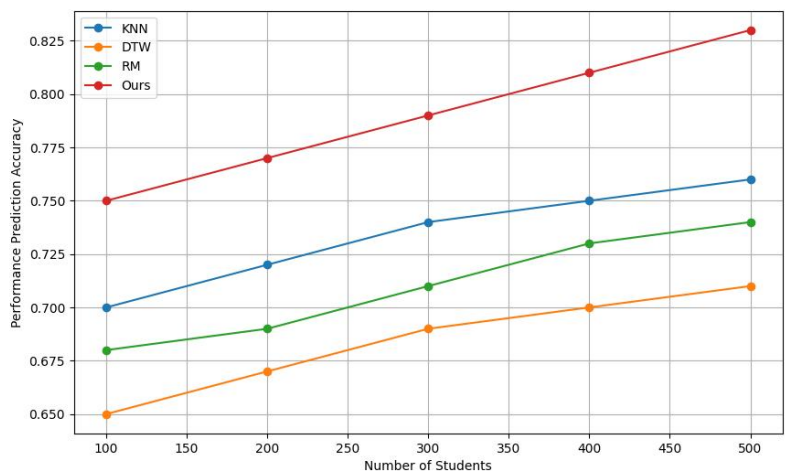

Figure 1: Accuracy Comparison of Different Methods.

The F1 score is a blended average of precision and recall, which is a combined reflection of these two metrics. The F1 score is especially important for unbalanced datasets because it provides a balance between different aspects of the model's performance. Following Figure 2 shows the F1 scores comparison results.

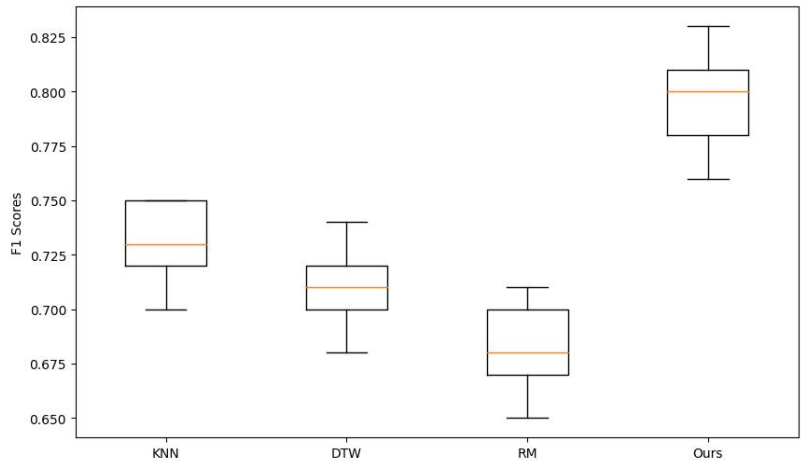

Figure 2: F1 Score Distribution Across Methods.

K-fold cross-validation is a technique for evaluating the generalization ability of a model. It divides the dataset into K subsets, then uses K-1 subsets to train the model, and uses the remaining subset to test the model. This process is repeated K times, each time a different subset is selected as the test set, and the final result is usually the average of the K evaluation results. This method can reduce the variation of model evaluation results caused by different data division methods, and make model evaluation more stable and reliable. Following Figure 3 shows K-fold cross-validation results.

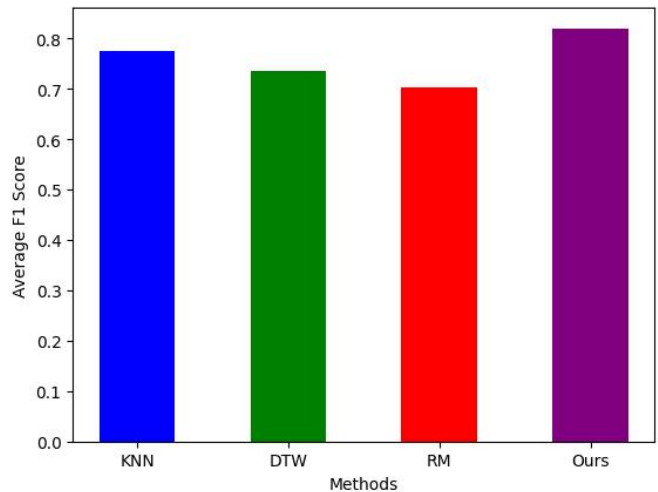

Figure 3: Comparison of F1 Scores Across Methods Using K-Fold Cross-Validation

## 5 CONCLUSION

In conclusion, the analysis of students' academic performance based on machine learning-based education big data shows that the use of multiple data science techniques and algorithms such as KNN, DTW, RM and our methods can effectively predict and improve student learning outcomes. Experimental results show that advanced machine learning models, such as ours, generally provide higher accuracy and F1 scores, especially in the case of K-fold cross-validation, where these models demonstrate superior stability and reliability. In addition, the application of different teaching methods and settings, such as traditional instruction, online learning platforms, and blended learning models, in data-driven decision support systems demonstrates the great potential of edtech solutions for personalized learning and student performance optimization. These findings highlight the importance of using big data and machine learning techniques in education to improve the quality and effectiveness of education, and also point to areas that need to be further explored in future research, including algorithm optimization, data privacy protection, and the integration of multimodal learning environments.